\documentclass[12pt]{article}
\usepackage{latexsym}

\def\K{{\cal K}}

\newcommand{\lbl}[1]{\label{eq: #1}}
\newcommand{\rf}[1]{\ref{eq: #1}}

\def\R{{\rm I\hspace{-.15em}R}}

\def\b{\begin{equation}}
\def\e{\end{equation}}
\def\bd{\begin{displaystyle}}
\def\ed{\end{displaystyle}}
\def\ba{\begin{array}}
\def\ea{\end{array}}

\def\bee{\begin{enumerate}}
\def\eee{\end{enumerate}}

\def\bes{\begin{eqnarray*}}
\def\ees{\end{eqnarray*}}
\def\be{\begin{eqnarray}}
\def\ee{\end{eqnarray}}

\begin{document}
\title{Linear Gravity in de Sitter Universe}
\author{ M.V.
Takook \thanks{e-mail: takook@razi.ac.ir} \thanks{Permanent address: Department of Physics, Razi University, Kermanshah, Iran }} \maketitle
\centerline{\it Groupe de physique des particules, D\'epartement
de physique,}\centerline{\it Universit\'e de Montr\'eal,
Montr\'eal, Qu\'ebec, Canada H3C 3J7}
\vspace{15pt}

\begin{abstract}
We give in this paper an explicit construction of the covariant
quantization of the rank-2 ``massless'' tensor field on de Sitter
(dS) space (linear covariant quantum gravity on a dS background).
The main ingredient of the construction is an indecomposable
representation of de Sitter group whith different
undecomposability channels. We here make the choice of a specific
gauge fixing in order to get the simplest possible structure of
the involved Gupta-Bleuler triplets. We describe the related Krein
space structure and  covariant field operators
$\K_{\alpha\beta}(x)$. We show that our gauge fixing  eliminates
any infrared divergence in the two-point function for the
traceless part of this field. But it is not possible to do the
same for the pure trace part (conformal sector). This work is in
the continuation of our previous ones concerning the ``minimally
coupled scalar fields  and the ``massive'' tensor field on dS.
\end{abstract}

\vspace{0.5cm} {\it Proposed PACS numbers}: 04.62.+v, 03.70+k,
11.10.Cd, 98.80.H \vspace{0.5cm}

\setcounter{equation}{0}
\section{Introduction}

The graviton propagator on dS space (in usual linear approximation
for the gravitational fields) for large separated points has a
pathological behaviour (infrared divergence) \cite{altu,
flilto,anmo1}. This behaviour may originate from imposing the
gauge invariance and then has no physical consequence. On the
other hand, some authors think infrared divergence could rather be
exploited in order to create instability of dS space \cite{for,
anilto1}. The field operator for linear gravity in dS space has
been considered along the latter line by Tsamis and Woodard in
terms of flat coordinates which cover only one-half of the dS
hyperboloid \cite{tswo}. They have examined the possibility of
quantum instability and they have found a quantum field which
breaks dS invariance. In this paper, we study the first
possibility, {\it i.e.} that quantization which respects gauge
invariance. We construct the covariant linear gravity field in de
Sitter space in terms of coordinates which cover the whole of the
dS hyperboloid .

In  previous papers we have considered the rank-2 ``massive''
tensor fields $\K(x)$ (divergenceless and traceless) on dS space
which carry  representations of the dS group belonging to the
principal series \cite{gata1}. We have shown that the principal
series and the representation which lies at the lower limit of the
discrete series $(p=2)$ have a precise physical meaning at the
flat space-time  limit $H=0$. On the other hand, for the discrete
series case, the tensor field $\K(x)$ obeying the divergenceless
condition $\partial.\K=0$ is singular. This type of singularity
appears precisely because of the divergenceless condition
$\partial.\K=0$. The latter is normally imposed in order to
associate an UIR's of the dS group to the ``massless'' tensor
field.

In this paper, we show that the above mentioned singularity in the
tensor field $\K(x)$ disappears when one gives up  the
divergenceless condition. This  implies that one has to enlarge
the space of states and to deal with an indecomposable
representation of the dS group. If we want to quantize $\K(x)$,
there appears another singularity in the Wightman two point
function like in the case of the ``massless'' minimally coupled
scalar fields \cite{alfo}. This type of singularity appears
because of the zero mode problem for the Laplace-Beltrami operator
on dS space. In order to eliminate it we must follow the procedure
already used in the quantization of the minimally coupled scalar
field \cite{gareta}.

The organization of this paper is the following. In section 2, we
present the de Sitter tensor field equation in terms of the
Casimir operator of the dS group and we describe its gauge
invariance properties. There are different levels of gauge
invariance and we lower the degree of gauge freedom by fixing the
values of gauge parameters.

In section 3 we describe what we mean by  minimal Gupta-Bleuler
triplet. We find that the central part of this triplet is, on the
representation level, an indecomposable representation of the dS
group.

Section 4 is devoted to the solutions of the field equation. We
construct these solutions in terms of a minimally coupled scalar
field $\phi$ and a projection operator ${\cal D}$ $$ \K (x)= {\cal
D}(x,\partial)\phi (x). $$ In the limit $H=0$ this becomes the
usual projection operator (the polarization tensor) for the
Minkowskian massless tensor field.  In section 5 we give an
explicit construction of the covariant two-point function and we
present three manners of constructing the field operator $\K(x)$.
Section 6 is a short conclusion and outlook.

\setcounter{equation}{0}
\section{de Sitter field equation}

Here, we briefly recall our de~Sitterian notations. The de Sitter
space-time is made identical to the four dimensional one-sheeted
hyperboloid
     \b X_H=\{x \in \R^5 ;x^2=\eta_{\alpha\beta} x^\alpha
 x^\beta =-H^{-2}\},\;\; \alpha,\beta=0,1,2,3,4, \e
where $\eta_{\alpha\beta}=$diag$(1,-1,-1,-1,-1)$. The de Sitter
metrics is \b  ds^2=\eta_{\alpha\beta}dx^{\alpha}dx^{\beta}=
g_{\mu\nu}^{dS}dX^{\mu}dX^{\nu},\;\; \mu=0,1,2,3,\e where $X^\mu$
are the 4 space-time coordinates in dS hyperboloid. Any
geometrical object can be written in terms of the four local
coordinates  $X^\mu$ or in terms of the five global coordinates
$x^\alpha$.

The wave equation for ``massless'' tensor fields $h_{\mu\nu}(X)$
propagating in de Sitter space can be derived through a
variational principle from the action integral \b\lbl{action}
S=-\frac{1}{16 \pi G}\int (R-2 \Lambda)\sqrt{-g}\, d^4X,\e where
$G$ is the Newtonian constant and  $\Lambda$ is the cosmological
constant. $\sqrt{-g}\,d^4X$ is the $O(1,4)$-invariant measure on
$X_H$. The variational calculus applied to (\rf{action}) leads to
the field equation \b R_{\mu\nu}-\frac{1}{2}Rg_{\mu\nu}+\Lambda
g_{\mu\nu}=0. \e We now expand the metric tensor as \b
g_{\mu\nu}=g^{dS}_{\mu\nu}+h_{\mu\nu},\e where $g^{dS}_{\mu\nu}$
is the gravitational background and $h_{\mu\nu}$ is the
fluctuation of the latter.

The wave equation which is obtained in the linear approximation
reads \cite{fr}: $$ -Lh \equiv
-(\Box_H+2H^2)h_{\mu\nu}-(\Box_H+H^2) g_{\mu\nu}h'-2
\nabla_{(\mu}\nabla^{\rho}h_{\nu)\rho}$$ \b\lbl{field-h}
+g_{\mu\nu}\nabla^{\lambda}\nabla^{\rho}h_{\lambda\rho}
+\nabla_{\mu}\nabla^{\nu}h'=0,\e where $\Box_H$ is the
Laplace-Beltrami operator on dS space and $h'=h_{\mu}^{\mu}$.
$\nabla^\nu$  is the covariant derivative on dS space. As usual,
two indices inside parentheses mean that they are symmetrized,
{\it i.e.} $T_{(\mu\nu)}=\frac{1}{2}(T_{\mu\nu}+T_{\nu\mu})$. The
field equation (\rf{field-h}) is invariant under the following
gauge transformation \b h_{\mu\nu} \longrightarrow
h_{\mu\nu}^{gt}=h_{\mu\nu}+\nabla_{(\mu}\Xi_{\nu)},\e where
$\Xi_{\nu}$ is an arbitrary vector field. One can choose a general
family of gauge conditions \b\lbl{gaugefix1} \nabla^{\mu}h_{\mu
\nu}=\zeta \nabla_{\nu}h',\e where $\zeta$ is a constant. We show
below that if we put $\zeta=\frac{1}{2}$, then the relation
between field and group representation is made quitely apparent.
This choice of gauge was first adopted by  Christensen and Duff
\cite{chdu}: \b\lbl{gaugefix2} \nabla^{\mu}h_{\mu
\nu}=\frac{1}{2}\nabla_{\nu}h'.\e

We shall now adopt the tensor field notation $\K_{\alpha\beta}(x)$
in order to make the relation between the field and the
representation of the dS group easier. In these notations, the
solutions to the field equations are written out very easily in
terms of scalar fields.  The tensor field $h_{\mu\nu}(X)$ is
locally determined by the tensor field $\K_{\alpha\beta}(x)$
through the relation \b\lbl{passage} h_{\mu\nu}(X)=\frac{\partial
x^{\alpha}}{\partial X^{\mu}}\frac{\partial x^{\beta}}{\partial
X^{\nu}}\K_{\alpha\beta}(x(X)),\;\; \K_{\alpha\beta}(x)
 =\frac{\partial X^{\nu}}{\partial x^{\alpha}}
\frac{\partial X^{\mu}}{\partial x^{\beta}}h_{\mu\nu}(X(x)). \e
The field $\K_{\alpha\beta}(x)$ lives on de Sitter space-time and
will be viewed here as an homogeneous function in the
$\R^5$-variables $x^{\alpha}$ with some arbitrarily chosen degree
$\sigma$ \b x^{\alpha}\frac{\partial }{\partial
x^{\alpha}}\K_{\beta\gamma}(x)=x\cdot\partial \K_{\beta\gamma}
(x)=\sigma \K_{\beta\gamma}(x). \e It also satisfies the
conditions of transversality \cite{di} \b x\cdot\K(x)=0,\mbox{ \it
i.e. }x^\alpha \K_{\alpha\beta}(x)=0,\mbox{ and } x^\beta
\K_{\alpha\beta}(x)=0 . \e

In order to obtain the wave equation for the tensor field $\K$, we
must use the tangential (or transverse) derivative $\bar \partial$
on de Sitter space \b \bar \partial_\alpha=\theta_{\alpha
\beta}\partial^\beta=
\partial_\alpha  +H^2x_\alpha x\cdot\partial,\;\;\;x\cdot\bar \partial=0,\e
where $\theta_{\alpha \beta}=\eta_{\alpha \beta}+H^2x_{\alpha}x_{
\beta}$ is the transverse projector. The algebraic machinery valid
for describing fields in anti-de Sitter space can be easily
transferred {\it mutatis mutenda} to dS space formalism $$
Q_s^{AdS}\longrightarrow -Q_s^{dS},  \;\;\;
(H^2)^{AdS}\longrightarrow -(H^2)^{dS}.$$ The field $\K$
corresponds to $h$ through (\rf{passage}) so we have \cite{ga} \b
\nabla_{\mu}h_{\nu\rho}\longrightarrow
\theta_{\alpha}^{\alpha'}\theta_{\beta}^{\beta'}
\theta_{\gamma}^{\gamma'}
\partial_{\alpha'}\K_{\beta'\gamma' },\e
Therefore one obtains the field equation for $\K$ from
(\rf{field-h}) \cite{fr} \b B[(Q_2+6)\K(x)+D_2\partial_2
\cdot\K]=0,\lbl{field-K} \e where $BT=T-\frac{1}{2}\theta T'$ with
$T':=\eta^{\alpha \beta} T_{\alpha \beta}$. The operator  $D_2$ is
the generalized gradient \b D_2K=H^{-2}{\cal S}(\bar
\partial-H^2x)K,\e
where ${\cal S}$ is the symmetrizer operator. The operator
$\partial_2 $ is the generalized divergence \b \partial_2\cdot
\K=\partial^T\cdot\K-\frac{1}{2}H^2D_1\K'=\partial \cdot\K- H^2 x
\K'-\frac{1}{2} \bar  \partial \K',\e where
$\partial^T\cdot\K=\partial.\K-H^2x\K'$ is the transverse
divergence. One can invert the operator $B$ and hence write the
equation (\rf{field-K}) in the form \b\lbl{field-K2}
(Q_2+6)\K(x)+D_2\partial_2 .\K=0.\e This equation is gauge
invariant, {\it i.e. } $\K^{gt}=\K+D_2\Lambda_g$ is a solution of
(\rf{field-K2}) for any vector field $\Lambda_g$ as far as $\K$
is.  The equation (\rf{field-K2}) can be derived from the
Lagrangian density \b\lbl{lagrangien} {\cal
L}=-\frac{1}{2x^2}\K..(Q_2+6)\K+(\partial_2 \cdot\K)^2.\e   The
gauge fixing condition (\rf{gaugefix1})
 reads in our notations
\b  \partial_2\cdot\K=(\zeta-\frac{1}{2})\bar \partial \K',\e
which becomes for $\zeta=1/2$, \b  \partial_2\cdot\K=0.\e

We first consider the latter case $\zeta=\frac{1}{2}$. Similarly
to the flat space QED, gauge fixing is accomplished by adding to
(\rf{lagrangien}) a gauge fixing term: \b {\cal
L}=-\frac{1}{2x^2}\K..(Q_2+6)\K+(\partial_2
\cdot\K)^2+\frac{1}{\alpha}  (\partial_2 \cdot\K)^2.\e The
variation of ${\cal L} $ then leads to the equation \cite{gaha} \b
(Q_2+6)\K(x)+cD_2\partial_2 \cdot\K=0.\e where
$c=\frac{1+\alpha}{\alpha}$ is a gauge fixing term. Actually, the
simplest choice of $c$ is not zero, as it will be shown later.

\noindent{\it Remark}: In the general gauge condition
(\rf{gaugefix1}) the gauge fixing Lagrangian reads \b {\cal
L}=-\frac{1}{2x^2}\K..(Q_2+6)\K+\frac{1}{2}(\partial_2
\cdot\K)^2+\frac{1}
{\alpha}(\partial_2\cdot\K-(\zeta-\frac{1}{2})\bar
\partial \K')^2.\e
The field equation which  derives from this reads
$$(Q_2+6)\K(x)+D_2\partial_2 \cdot\K
+\frac{1}{\alpha}[D_2\partial_2 \cdot\K+(\zeta-\frac{1}{2})^2\eta
(\bar \partial)^2\K'-(\zeta-\frac{1}{2})(D_2\bar \partial
\K'-{\cal S}\bar \partial\partial_2 \cdot\K)] =0. $$ In the
following we shall work with the choice $\zeta=\frac{1}{2}$  only.

\setcounter{equation}{0}
\section{Gupta-Bleuler triplet}

>From now on we shall deal with the field that is the solution of
the equation $(2.23)$ \b (Q_2+6)\K(x)+cD_2\partial_2 .\K=0.\e
Because of the algebraic machinery valid for describing fields in
anti-de Sitter space, the following dS invariant bilinear form (or
inner product) on the space of solutions is define as
\cite{gahamu}
\renewcommand{\arraystretch}{0.6}
\begin{equation}
(\K_1,\K_2)=\frac{i}{H^2}\int_{\renewcommand{\arraystretch}{0.4}
\!\!\begin{array}{l}
\begin{scriptstyle}S^3\end{scriptstyle}\\ \begin{scriptstyle}\rho=0
\end{scriptstyle}\end{array}}\!\!\! \K_{1\alpha\beta}^*(\rho,\Omega).A_c
\K_2^{\alpha\beta}(\rho,\Omega)d\Omega,\end{equation}
\renewcommand{\arraystretch}{1}
where we have used the (conformal) coordinate system defined in
$(4.18)$ below. $A_c$ is a certain $c$-dependent matrix
differential operator, which their form can be calculated using
the field equation $(3.1)$. For fields that satisfy the
divergenceless condition, $A_c$ is equal to
$\overrightarrow{\partial}_{\rho}-\overleftarrow{\partial}_{\rho}$
(independent of $c$). In the next, we shall always mean by
solution to (3.1) a field which is square integrable with respect
to (3.2). Let us now define the Gupta-Bleuler triplet of carrier
spaces in order to build a covariant quantum  field \cite{ga}.
When $c=1$, the gauge field $\K_g=D_2\Lambda_g$ is the solution of
the free-field equation $(3.1)$ where $\Lambda_g$ is an arbitrary
vector field. With another choice of value $c\neq 1$, the gauge
field $K_g=D_2\Lambda_g$ is solution to the field equation only if
$\Lambda_g$ obeys \b (Q_1+6)\Lambda_g =0,\;\;x. \Lambda_g=0 .\e
The divergence $\partial_2.\K$ is a vector field and if $\K$  is a
general solution to Eq. (3.1), then $\partial_2.\K$ is also a
solution to \b (Q_1+6)\partial_2.\K =0.\e The space of solutions
of $(3.3)$ and $(3.4)$ coupled to the divergenceless condition
carries a nonunitary irreducible representation
$U^{(1,\nu=i\frac{5}{2})}$, which we can call ``imaginary
massive'' representation of the dS group.

We shall denote by $V'$ the space of solutions to the field
equation $(3.1)$. We define the subspace $V \subset V'$ of tensor
fields satisfying $\partial.\K=0$, and $
(Q_2+6)\K=\Box_H\K_{\alpha\beta}(x)=0$. Within this subspace there
exists a subspace $V_g$ of gauge fields defined by \b
\K=D_2\Lambda_g, \;\; (Q_1+6)\Lambda_g
=0,\;\;||D_2\Lambda_g||=0,\e where the norm is  (abusively)
defined by $(3.2)$. We thus see a chain appear \b V_g \subset V
\subset V' .\e The form $(3.2)$ is indefinite in the space $V'$
and is positive semidefinite in the subspace $V$. It becomes
positive definite on the coset space $V/V_g$ . The latter carries
the physical content of the theory and carries the unitary
representation $ \Pi^+_{2,2} \bigoplus \Pi^-_{2,2}$ in the Dixmier
notations \cite{dix}. The space $V_g$ carries the representation
$U^{(1,\nu=i\frac{5}{2})}$ associated with the equation $(3.3)$.
Also the coset space $V'/V$ carries the representation
$U^{(1,\nu=i\frac{5}{2})}$ associated with the equation $(3.4)$.
Now, if we retain in the whole space $V'$ those solutions which
obey the following extra-conditions $\partial.\partial_2
.\K=0,\partial .D_2\K=0$,
 we obtain a set which carries the indecomposable representation
\b U^{(1,\nu=i\frac{5}{2})} \longrightarrow \{\Pi^+_{2,2}
\bigoplus \Pi^-_{2,2}\} \longrightarrow U^{(1,\nu=i\frac{5}{2})},
\e where the arrow means a semidirect sum. Space $V$ does not
depend on $c$. On the contrary the space $V'$ depends on $c$. The
value $c=\frac{2}{5}$ is remarkable for it corresponds to the
simplest case, {\it i.e. } to the minimal structure. Alone the
central part is physical. It describes fields which obey the
equations \b     \Box_H\K_{\alpha\beta}(x)=0 , \;\; \mbox{and}
\;\;\partial. \K=0 .\e We should here note the appearance of the
additional global gauge freedom due to the homogeneity of the
above equations: \b \K\longrightarrow \K^{gt}=\K+\mbox{const.}.\e
However, one cannot construct a constant tensor field that satisfy
the condition $x.\K^{gt}=0$ {\it i.e.}, and so this constant
tensor field has no physical meaning. The zero mode singularity
which results from it is of the same type as that one which
appears in the case of the massless minimally coupled scalar
field, and its elimination requires the same techniques as those
ones which were employed in \cite{gareta}. That means one must to
enlarge the coset space $V/V_g$ (`` central parter'') to include
the negative norme state and one it replace also with an
indecomposable representation.

\setcounter{equation}{0}
\section{dS-field solution}

A general solution of Equation $(3.1)$ can be constructed by a
scalar field and two vector fields. Let us introduce a tensor
field $\K$ in terms of a five-dimensional constant vector
$Z=(Z_\alpha)$ and a scalar field $\phi_1$ and two vector fields
$K$ and $K_g$ by putting \b \K=\theta\phi_1+ {\cal S}\bar
Z_1K+D_2K_g,\e where $\bar Z_{1\alpha}=\theta_{\alpha\beta}
Z^{\beta}_1$. Substituing  $\K$ into $(3.1)$ and make use of the
commutation rules and algebraic identities for the various
involved operators and fields \cite{gaha}, \b Q_2\theta
\phi=\theta Q_0\phi,\e \b  Q_2 {\cal S}\bar Z K={\cal S}\bar Z
(Q_1-4)K-2H^2D_2x.ZK+4\theta Z.K .\e \b  Q_2D_2K_g=D_2Q_1K_g,\e \b
\partial_2 .\theta\phi=-H^2D_1\phi,  \e \b \partial_2 . {\cal
S}\bar Z_1K=\bar Z_1 \partial_1 . K-H^2D_1Z.K-H^2xZ.K+Z.(\bar
\partial+5H^2x)K, \e \b \partial_2 . D_2K_g = -(Q_1+6)K_g,\e
we find that $K$ obeys the wave equation $$( Q_1+2)K+cD_1\partial
.K=0,\; x.K=0,$$ where $D_1=H^{-2}\bar \partial$. If we impose the
supplementary condition $\partial.K=0$, then we get \b (
Q_1+2)K=0,\; x.K=0=\partial.K,\e the solutions have a dS group
representation interpretation. Now, if we make the choice \b
\phi_1 =-\frac{2}{3}Z_1.K, \e then we obtain the equation
 \b Q_0 \phi_1=0 .\e
Hence, $\phi_1$ is a ``massless'' minimally coupled scalar field.
The other field in ($4.1$), $K_g$, is also determined in terms of
$K$ $$ (Q_1+6)K_g =\frac{c}{2(c-1)}H^2 D _1\phi_1 +
\frac{2-5c}{1-c}  H^2 x. Z_1 K+$$ \b \frac{c}{1-c} (H^2 x Z_1.K -
Z_1.\bar\partial K).\e If we chose $c=\frac{2}{5}$, then we get
the simplest form for $K_g$. \b  K_g =\frac{1}{9}\left(H^2 x
Z_1.K- Z_1.\bar\partial K+\frac{2}{3}H^2 D_1 Z_1.K\right).\e In
conclusion, if we know the vector field $K$, we also know the
tensor field $\K$. In order to determine the latter, we proceed in
the same manner as in the above. We choose the folowing form for
the vector field $K$ solution to ($4.8$) \b K_{\alpha}=\bar
Z_{2\alpha}\phi_2+D_{1\alpha}\phi_3.\e Affter putting in $(4.8)$,
we find for $\phi_2$ and $\phi_3$, $$  Q_0\phi_2=0.$$ \b \phi_3
=-\frac{1}{2} [Z_2.\bar \partial \phi_2+2H^2 x.Z_2\phi_2].\e
$\phi_2$ is also a ``massless'' minimally coupled scalar field.
Therefore we can construct the tensor field $\K$ in terms of two
``massless'' minimally coupled scalar fields $\phi_1$ and
$\phi_2$. But both fields are related by $(4.9)$. Therefore one
can write \b \K_{\alpha \beta}(x)={\cal D}_{\alpha
\beta}(x,\partial,Z_1,Z_2)\phi,\;\; \phi=\phi_2,\e where $$ {\cal
D}(x,\partial,Z_1,Z_2)=\left(-\frac{2}{3}\theta Z_1.+{\cal S}\bar
Z_1+\frac{1}{9 }D_2 (H^2 xZ_1.-Z_1.\bar \partial +\frac{2}{3}H^2
D_1 Z_1.)\right)$$ \b\left( \bar Z_{2}-\frac{1}{2}  D_{1}(Z_2.\bar
\partial+2H^2x.Z_2)\right),\e
and $\phi$ is a ``massless'' minimally coupled scalar field with
homogeneous degree $\sigma=0$ or $-3$. It is given in the form of
a dS plane wave \b \phi(x)=(Hx.\xi)^\sigma ,\e where $\xi \in \R^5
$ lies on the null cone $ {\cal C} = \{ \xi \in \R^5;\;\; \xi^2=0
\}$. Due to the zero mode problem ($\sigma=0)$, one cannot
construct a covariant quantum field. For restoring covariance, we
make the solution explicit in a system of bounded global
coordinates $(X^\mu,\;\mu=0,1,2,3)$ well-suited to describing a
compactified version of dS, namely S$^3 \times{\rm S}^1$ (Lie
sphere). This system is given by \b \left\{ \ba{rcl}
x^0&=&H^{-1}\tan \rho\\
x^1&=&(H\cos\rho)^{-1}\,(\sin\alpha\,\sin\theta\,\cos\phi),\\
x^2&=&(H\cos\rho)^{-1}\,(\sin\alpha\,\sin\theta\,\sin\phi),\\
x^3&=&(H\cos\rho)^{-1}\,(\sin\alpha\,\cos\theta),\\
x^4&=&(H\cos\rho)^{-1}\,(\cos\alpha),\ea\right.\e where
$-\pi/2<\rho<\pi/2$, $0\leq\alpha\leq\pi$, $0\leq\theta\leq\pi$
and $0\leq\phi \leq 2\pi$. The closure of the $\rho$-interval is
actually involved when dealing with conformal action on
compactified space-time. The de Sitter metrics now reads \b
ds^2=g^{dS}_{\mu\nu}dX^\mu dX^\nu=\frac{1}{H^2 \cos^2\rho}
(d\rho^2-d\alpha^2-\sin^2\alpha\,
d\theta^2-\sin^2\alpha\sin^2\theta \,d\phi^2).\e The solution to
the field equation $$Q_0\phi(x)=0=\Box_H\phi(x),$$
 reads in this coordinate system
\cite{gareta}: \b\phi_{Llm}(x)=X_{L}(\rho)Y_{Llm}(\Omega),\e with
\b X_{L}(\rho)=\frac{H}{2}[2(L+2)(L+1)L]^{-\frac{1}{2}} \left(L
e^{-i(L+2)\rho}+(L+2)e^{-iL\rho}\right),\e for $L\neq 0$ and
$$\phi_{000}=\phi_g+\frac{1}{2}\phi_s,\;\phi_g=\frac{H}{2\pi}\;\;
\phi_s=-i\frac{H}{2\pi}[\rho+\frac{1}{2}\sin 2\rho ],$$ for $L=0$.
The $y_{Llm}(\Omega)$'s are the hyperspherical harmonics. The
solution $(4.15)$ can be written as \b \K_{\alpha \beta}(x)={\cal
D}_{\alpha \beta}(x,\partial,Z_1,Z_2)\phi_{Llm}(x)={\cal
D}_{\alpha \beta}(x,\partial,Z_1,Z_2)X_L(\rho)y_{Llm}(\Omega).\e
$Z_1$ and $Z_2$ are two constant vectors. We choose them in such a
way that in the limit $H=0$, one obtains the polarization tensor
in the Minkowskian space \b \lim_{H \rightarrow 0}{\cal D}_{\alpha
\beta}(x,\partial,Z_1,Z_2)\frac{X_L(\rho)y_{Llm}(\Omega)}{H\sqrt
H} \equiv \epsilon_{\mu\nu}(k)\frac{e^{ ik.X}}{\sqrt{k_0}} ,\e
where $\epsilon_{\mu\nu}(k)$ is the polarization tensor in the
Minkowski space-time \cite{we}:
$$k^{\mu}\epsilon_{\mu\nu}(k)-\frac{1}{2}k_{\nu}\epsilon_{\nu}^{\nu}(k)=0,$$

\b
\epsilon_{\mu\nu}(k)=\epsilon_{\nu\mu}(k),\;\;k^{\nu}k_{\nu}=0.\e
Finally, we write the solution under the form \b \K_{\alpha
\beta}(x)={\cal
D}_{\alpha\beta}^{\lambda}(x,\partial)\phi_{Llm}(\rho,\Omega)\equiv
{\cal
E}_{\alpha\beta}^{\lambda}(\rho,\Omega,Llm)\phi_{Llm}(\rho,\Omega),\e
where ${\cal E}$ is the generalized polarization tensor and the
index $\lambda$ runs on all possible polarizations. The explicit
form of the polarization tensor is actually not important here.
Indeed,  one can find the two-point function by just using the
recurrence formula $(4.1)$. On the other hand, one can calculate
the projection operator ${\cal D}$ in the coordinate system
$(4.18)$ and makes it act on the scalar field $(4.20)$. Then one
considers the behavior at the limit $H=0$ and compares with the
polarization tensor $(4.24)$. This is the way we determine the
generalized polarization tensor ${\cal E}$.

The solution ($4.22$) is traceless $\K'=0$. Let us now consider
the pure trace solution (conformal sector) \b
\K^{pt}=\frac{1}{4}\theta \psi. \e If we report in $(3.1)$, we
obtain $$  (Q_0+6)\psi+\frac{c}{2}Q_0\psi=0,$$ or \b
(Q_0+\frac{12}{2+c})\psi=0.\e On the other hand, any scalar field
corresponding to the discrete series representation of the dS
group obeys the equation \b  (Q_0+n(n+3))\psi=0.\e Hence we see
that the value  $c=\frac{2}{5} $ does not correspond to a unitary
irreducible representation of the dS group. But there exists a
nonunitary representation corresponding to that $c=\frac{2}{5} $
$$ (\Box_H-5H^2)\psi=0.$$ Difficulties arise when we want to
quantize these fields with the negative mass square (conformal
sector with $c> -2$ and discrete series with $n>0$) . The
two-point functions for these fields have a pathological
large-distance behavior. If we choose  $c < -2$ this pathological
behavior for the conformal sector disappears but it is still
present in the traceless part.

We just stress on the fact that, so far, this degree of freedom
does not appear as a physical one. In  the next section,  we shall
consider only the traceless part. We shall deal with the conformal
sector in the next paper.

\setcounter{equation}{0}
\section{Two-point function and Field operator}

In a previous paper devoted to the ``massive'' spin-$2$ field
(divergenceless or `` transverse'' and traceless), we have
constructed the quantum field from the Wightman two-point function
${\cal W}$. The latter is defined by \cite{gata1} \b {\cal
W}^\nu_{\alpha\beta \alpha'\beta'}(x,x')=\langle
\Omega,\K_{\alpha\beta}(x)\K_{\alpha'\beta'}(x')\Omega  \rangle ,
\;\;\alpha,\beta, \alpha',\beta'=0,1,..,4.\e where $x,x'\in X_H$
and $\mid \Omega  \rangle$ is the Fock vacuum state. We have found
that this function can be written under the form \b {\cal
W}^\nu_{\alpha\beta \alpha'\beta'}(x,x')=D_{\alpha\beta
\alpha'\beta'}(x,x'){\cal W}^\nu(x,x'), \e where ${\cal
W}^\nu(x,x')$ is the Wightman two-point function for the massive
scalar field and $D_{\alpha\beta \alpha'\beta'}(x,x')$ is the
projection tensor. Of course, we crudely could replace $\nu$
(principal-series parameter) by $\pm \frac{3}{2}i$
(discrete-series parameter) in order to get the ``massless''
tensor field associated to linear quantum gravity in dS space.
However this procedure make two types of singularity appear in the
definition of the Wightman two-point function. The first one
appears in the projection tensor $D_{\alpha\beta
\alpha'\beta'}(x,x')$ and it disappears if one fixes the gauge
$(c=\frac{2}{5})$. The other one appears in the scalar Wightman
two-point function ${\cal W}(x,x')$ which corresponds to the
minimally coupled scalar field. In order to manage the latter type
of singularity, one has the choice between three procedures
already present in the minimally coupled scalar field case.

For the first one, after fixing the gauge, in the same way that we
have done for the ``massive'' tensor field, using the equation
$(4.16)$ the tensor two-point function can be written in the form
\b {\cal W}_{\alpha\beta \alpha'\beta'}(x,x')=\Delta_{\alpha\beta
\alpha'\beta'} (x,x'){\cal W}(x,x'), \e where $$
\Delta(x,\partial,x',\partial')=\left(-\frac{2}{3}\theta
\theta'+\frac{H^2}{9}D_2D_2' [ H^2xx' +\frac{2}{3}H^2 D_1
D'_1]\right)$$ $$\left(\theta..\theta' +\frac{1}{2H^2}\bar
\partial.\bar
\partial'[H^{-2}\bar \partial.\bar \partial'+2H^2
x.x']\right)+$$ \b \left({\cal S}{\cal
S}'\theta.\theta'+\frac{1}{9}D_2D_2'\bar
\partial.\bar\partial'  \right)
 \left(\theta.\theta'-\frac{1}{2H^2}\partial
\partial'[H^{-2}\bar \partial.\bar \partial'+2H^2
x.x']\right),\e and ${\cal W}$ is the two-point function for the
minimally coupled scalar field. Now the the first type of
singularity disappear.

Now we consider the second type of singularity. In the first
procedure, one defines the Fock-vacuum state and keeps the
condition of positivity for the ``central parte'', but one
abandons the de Sitter-invariance \cite{alfo}. In the second
procedure, one keeps the de Sitter-invariance, but one defines a
ground state which is \underline{not} a Fock-vacuum state
\cite{kiga}. In the third procedure, the full de Sitter-invariance
is preserved. For that, we define a vacuum state in the framework
of Krein spaces and Gupta-Bleuler quantization (``Gupta-Bleuler
vacuum''), but the price to pay is then to give up the positivity
condition for the ``central parte'' \cite{gareta}. Let us now
consider briefly these three cases.

The first is the way adopted by Allen and Folacci. In this case,
the field operator is defined by \b \K_{\alpha
\beta}(x)=\sum_{\lambda,L}a_{Llm}^{\lambda}{\cal D}_{\alpha
\beta}^{\lambda}(x,\partial) X_L(\rho)y_{Llm}(\Omega)+H.C.,\e
where $X_0(\rho)=\phi_g+\phi_s$ and with the vacuum condition
 \b   a_{Llm}^{\lambda}|O>_{A,B}=0, \;\;\forall \;\;\;0\leq l \leq L, \;\;\; -l\leq m \leq l,\e
$$[a_{Llm}^{\lambda},a_{L'l'm'}^{ \lambda '}]=\delta_{\lambda
\lambda '}\delta_{LL'}\delta_{ll'}\delta_{mm'}.$$ Then the
two-point function reads \b {\cal W}_{\alpha\beta
\alpha'\beta'}(x,x')=\Delta_{\alpha\beta \alpha'\beta'}(x,x'){\cal
W}_p(x,x'), \e where ${\cal W}_p$ is the two-point function for
the minimally coupled scalar field in the $O(4)$-vacuum state
\cite{fo} \b {\cal W}_p(x,x')=\frac{H^2}{8\pi^2}[\frac{1}{1-{\cal
Z}}-\ln (1-{\cal Z})+\ln 2+f_{AB}(\rho,\rho')], \e
$f_{AB}(\rho,\rho')$ is a function of the variables $\rho$ and
$\rho'$ which breaks dS-invariance. In order to obtain the
two-point function $(5.7)$ from the field operator $(5.5)$, the
projection operator ${\cal D}$ has to satisfy the relation $$
\sum_{\lambda} {\cal D}_{\alpha \beta}^{\lambda}(x,\partial) {\cal
D}_{\alpha' \beta'}^{\lambda}(x',\partial')=\Delta_{\alpha\beta
\alpha'\beta'}(x,\partial;x',\partial').$$

The second case corresponds to the so-called naive dS-invariance,
which is restored at the cost of  giving up the existence of the
Fock vacuum state. The alternative is to define a ground state
where is not normalisable. In this case one cannot define the
Wightman two-point functiondue to its singular behavior. The field
operator is defined by $$ \K_{\alpha \beta}(x)=\sum_{\lambda,L\neq
0} a_{Llm}^{\lambda}{\cal D}_{\alpha \beta}^{\lambda}(x,\partial)
X_L(\theta)y_{Llm}(\Omega)+H.C.$$
 \b +\frac{H}{\sqrt 2}\sum_{\lambda}[Q^{\lambda}
  {\cal  D}_{\alpha \beta}^{\lambda}(x,\partial)\phi_g+
{\cal D}_{\alpha \beta}^{\lambda}
(x,\partial)(\frac{1}{2}\sin2\rho + \rho )        P^{\lambda}],\e
where $$[Q^{\lambda},P^{\lambda'}]=i\delta_{\lambda\lambda'},
\;\;[a_{Llm}^{\lambda},a_{L'l'm'}^{ \lambda '}]=\delta_{\lambda
\lambda '}\delta_{LL'}\delta_{ll'}\delta_{mm'},$$ $$
a_{Llm}^{\lambda}|O>=0,\;\; L \geq 1,$$
 \b P^{\lambda}|O>=0,\;\; L=0.\e

In the third case we add the negative norm states
 $L \leq -2$ or equivalent to the negative commutation relation field.  The corresponding field operator is defined by
$$ \K_{\alpha \beta}(x)=\sum_{\lambda,L} a_{Llm}^{\lambda}{\cal
D}_{\alpha \beta}^{\lambda}(x,\partial)
X_L(\rho)y_{Llm}(\Omega)+H.C.$$ \b +\sum_{\lambda,L}
b_{Llm}^{\lambda}{\cal D}_{\alpha \beta}^{\lambda}(x,\partial)
X^*_L(\rho)y_{Llm}(\Omega)+H.C.,\e where $a$ and $b$ are defined
by \b    a_{Llm}^{\lambda}|0>=0,\;\; b_{Llm}^{\lambda}|0>=0,
\;\;\forall \;\;\;0\leq l \leq L, \;\;\; -l\leq m \leq l,\e
\b[a_{Llm}^{\lambda},a_{L'l'm'}^{ \lambda '}]=\delta_{\lambda
\lambda '}\delta_{LL'}\delta_{ll'}\delta_{mm'},\e
\b[b_{Llm}^{\lambda},b_{L'l'm'}^{ \lambda '}]=-\delta_{\lambda
\lambda '}\delta_{LL'}\delta_{ll'}\delta_{mm'}.\e Then the
two-point function reads \b {\cal W}_{\alpha\beta
\alpha'\beta'}(x,x')=\Delta_{\alpha\beta \alpha'\beta'}(x,x'){\cal
W}(x,x'), \e where ${\cal W}$ is the two-point function for the
minimally coupled scalar field in the ``Gupta-Bleuler vacuum''
state \cite{gata2} \b {\cal W}(x,x')=\frac{iH^2}{8\pi^2} \epsilon
(x^0-x'^0)[\delta(1-{\cal Z}(x,x'))+\vartheta ({\cal Z}(x,x')-1)],
\e with \b \epsilon (x^0-x'^0)=\left\{ \ba{rcl} 1&x^0>x'^0\\
0&x^0=x'^0\\ -1&x^0<x'^0.\\ \ea\right.\e The function (5.15)  does
not satisfy the condition of positivity for the ``central parte''.
If we wish that the positivity condition holds true, we must
eliminate the negative norm state $(b^\dagger)^n|0>$ then we break
dS invariance and one obtaie the Allen procedure. Let insist here
that the third procedure only allows us to avoid the pathological
large-distance behavior of the physical graviton propagator.

\section{Conclusion and outlook}

In this paper we have presented three ways of constructing a
quantum linear gravity (traceless part) on dS space. Only in the
third one we have been able to get a covariant two-point function,
which is free of the pathological large-distance behavior.
Antoniadis, Iliopoulos and Tomaras have also  shown that the
pathological large-distance behavior of the graviton propagator on
a dS background does not manifest itself in the quadratic part of
the effective action in the one loop approximation \cite{anilto2}.
That means that this behaviour may be gauge dependent and it
should not appear in an effective way in a physical quantity. On
the other hand, it exists in an irreducible way in the pure-trace
part (conformal sector).

In the next paper we shall consider the linear conformal gravity
and also the conformal sector in dS space. The former is important
for understanding the flat space-time limit ($H=0$). The later is
interesting for inflationary universe scenarii. In these theories,
one introduces an inflaton scalar field. Because of this field,
the conformal sector of the metric becomes dynamical and it must
be quantized \cite{anmo2, anmamo}. Then it produces a
gravitational instability. This gravitational instability and the
primordial quantum fluctuation of the inflaton scalar field define
the inflationary model. The latter can explain the formation of
the galaxies, clusters of galaxies and the large scale structure
of the universe \cite{lepost} .

\noindent {\bf{Acknowlegements}}: We are grateful to J.
Iliopoulos, J.P. Gazeau, S. Rouhani and  M.B. Paranjape for
helpful discussions.

\end{document}